\documentstyle[pre,aps,epsf,psfig]{revtex}

\newcommand \be{\begin{equation}}
\newcommand \ee{\end{equation}}
\newcommand \ba{\begin{eqnarray}}
\newcommand \ea{\end{eqnarray}}

\newcommand \om {\omega}

\newcommand{\DK}[1]{\mbox{\boldmath$#1$}}
\newcommand{\BE}{\begin{equation}\label}
\newcommand{\BEQ}{\begin{eqnarray}\label}
\newcommand{\EE}{\end{equation}}
\newcommand{\EEQ}{\end{eqnarray}}

\begin{document}
\setcounter{page}{1}
\draft
\title{Semiclassical dynamics and time correlations in two-component plasmas
 }

\author{J.~Ortner $^{a)}$, I. Valuev $^{b)}$, and W.~Ebeling$^{a)}$}
\address{
$^{a)}${\it Institut f\"ur Physik, Humboldt Universit\"{a}t zu Berlin,\\ 
Invalidenstr. 110, D-10115 Berlin, Germany}\\
$^{b)}${\it {Department of Molecular and Chemical Physics, Moscow Institute of Physics\\ and Technology, 141700 Dolgoprudny, Russia}}
}
\date{\today}
\maketitle

\begin{abstract}
The semiclassical dynamics of a charged particle moving in a two-component plasma is considered using a corrected Kelbg pseudopotential. 
We employ the  classical Nevanlinna-type theory of frequency moments to determine the  velocity and force autocorrelation functions. The constructed expressions preserve the exact short and long-time behavior of the autocorrelators. The short-time behavior is characterized by two parameters which are expressable through the plasma static correlation functions. The long-time behavior is determined by the self-diffusion coefficient. The theoretical predictions are compared with the results of semiclassical molecular dynamics simulation.
\end{abstract}


\section{Introduction}

The purpose of this paper is the investigation of the dynamics of force on a charged particle in a two component plasma. Boercker {\em et al.} have shown the effect of ion motion on the spectral line broadening by the surrounding plasma \cite{BID87,B93}. In recent papers it was argued that the microfield dynamics influence the fusion rates \cite{RomEb98} and rates for three-body electron-ion recombination \cite{Rom98} in dense plasmas. Generally speaking, to calculate the plasma effect on rates and spectral line broadening one needs a theory of average forces and microfields, including the resolution in space and time. Basic results in this field were obtained by Silin and Rukhadze \cite{SR64}, Klimontovitch \cite{K82}, Alastuey et. al. \cite{ALL91}, and Berkovsky et. al. \cite{BDCST96}.

The determination of the static distribution of the ion or electron component of the electric microfield is a well studied problem (for a review see \cite{dufty}). The corresponding investigations are performed on the basis of the one-component plasma (OCP) model. A straightforward generalization of the OCP model is the model of a two-component plasma (TCP), consisting of electrons and ions. In a recent paper \cite{OVE99} the probability distribution for the electric microfield at a charged point has been studied. It was shown that the two-component plasma microfield distribution shows a larger probability of high microfield values than the corresponding distribution of the OCP model.

The dynamics of the electric microfield is a less understood problem than that of the static microfield distribution even for the case of an OCP. Recently some progress has been made for both the case of electric field dynamics at a neutral point \cite{ALL91,dufty-zogaib,BDCST95} and the dynamics of force on a charged impurity ion in an OCP \cite{BDCST96}.

This paper is aimed to extend the studies of electric microfield dynamics in OCP to the case of an equilibrium two component plasma. For simplicity we consider a two-component plasma which is anti-symmetrical with respect to 
the charges ($e_- = - e_+$) and therefore symmetrical with respect to the densities
($n_i = n_e$). 
To simplify the numeric investigations we simulated a mass symmetric (nonrelativistic) electron-positron plasma with $m= m_i= m_e$. The theoretical investigations are carried out for arbitrary electron-ion mass ratios. \\

$^1$ Dedicated to the $75$th birthday of Youri L. Klimontovich

\newpage

In this paper we will study the dependence of the force dynamics on the coupling constant $\Gamma=e^2/k_BTa$ of the plasma, where $T$ is the temperature, and $a=(3/4 \pi n_e)^{1/3}$ is the average distance between the electrons.  Coupled plasmas with a plasma parameter of the order or greater than unity are important objects
in nature, laboratory experiments, and in 
technology \cite{gruenes-buch,braunes-buch,ichimaru92,ksruegen}.
Recent lasers allow to create a coupled plasma within femtoseconds \cite{THWS96}. Laser generated plasmas are nonequlibrium plasmas with an initial electron temperature much higher than the ion temperature. However, in this paper we restrict our considerations to the model object of an equilibrium two-component plasma (TCP). 

Several investigations were
devoted to the simulation of equilibrium two-component plasmas
Being interested in quasi-classical methods we mention explicitely the quasi-classical 
simulations of two-component plasmas performed by Norman and by Hansen 
\cite{norman,hansen2}. 

In this paper the free charges (electron and ions) are simulated by a semi-classical dynamics based on effective potentials.  
The idea of the semi-classical method  is to incorporate
quantum-mechanical effects (in particular the
Heisenberg and the Pauli principle) by appropriate potentials. This method was pioneered by Kelbg, Deutsch and others \cite{Kelbg,Deutsch}. 
Certainly, such a quasi-classical approach has several limits. For the calculation of a standard
macroscopic property as the microfield dynamics which has a well defined classical limit the semi-classical approach may be very useful. The advantage of such an approach is the relative simplicity of the algorithm.



\section{The Slater sum and the semiclassical model} \label{model}

A familiar derivation of effective potentials describing quantum effects is based on the Slater sums which are defined by the N - particle wave functions,

\be \label{SlaterN}
S( \DK{r}_1,\ldots,\DK{r}_N ) = \mbox{const} \sum \exp \left(-\beta\;E_n \right)\;
\left|\Psi_n\left( \DK{r}_1,\ldots,\DK{r}_N \right)\right|^2 \quad ,
\ee
where $E_n$ and $\Psi_n$  are the energy levels and corresponding 
wave functions  of the ensemble of $N$ particles with 
coordinates ${r}_1,\ldots,{r}_N$. Here we consider a two-component plasma consisting of $N_e$ electrons with mass $m_e$ and $N_i=N_e$ ions with mass $m_i$. 
The properties of the Slater 
sums for Coulombic systems
were studied in detail by several authors \cite{gruenes-buch,ebeling}. 
 Choosing the effective potential

\be
U^{(N)}(\DK{r}_1,\ldots,\DK{r}_N ) = - k_B T \ln S( \DK{r}_1,\ldots,\DK{r}_N ) \quad .
\label{slater}
\ee
we may calculate the correct thermodynamic functions of the original quantum system \cite{gruenes-buch,ebeling,norman} from the thermodynamic functions of a classical reference system.

The Slater sum may be considered as an analogue of the classical Boltzmann factor. Therefore it is straightforward to use the Slater sum for the definition of an effective potential. The only modification in comparison with classical theory is the appearance of many-particle interactions. If the system is not to dense (i.e., $n_e \Lambda_e^3 \ll 1$, $\Lambda_e = \hbar/\sqrt{2 m_{e}k_BT}$) one may neglect the contributions of higher order many-particle interactions. In this case one writes approximately,

\BE{pair}
U^{(N)}(\DK{r}_1,\ldots,\DK{r}_N ) \approx \sum_{i<j} u_{ij}(\DK{r}_i,\DK{r}_j) \quad,
\EE
where the effective two-particle potential $u_{ab}$ is defined by the two-particle Slater sum,
\BE{slater-sum}
 S_{ab}^{(2)} (r) = \exp \left( -\beta u_{ab}(r) \right)
   = {\rm const.} {\sum_{\alpha}}' \exp\left( -\beta E_\alpha \right) 
   \mid \Psi_\alpha \mid^2 \;\;.
\EE

Here $\Psi_\alpha$ and $E_\alpha$ denote the wave functions and energy
levels of the pair $ab$, respectively. The prime at the summation sign indicates that the contribution of the bound states (which is not be considered here) has to be omitted.
 
Principal it is possible to calculate the Slater sum for a pair of particles directly from the known two-particle Coulomb wavefunctions. To simplify the simulations it is better to have an analytic expression for the potential. A possible candidate is the so called Kelbg potential obtained by a perturbational expansion 
It reads \cite{Kelbg}
\begin{equation}\label{Kelbg-pot}
u_{ab}(r)={e_ae_b\over{r}}F(r/\lambda_{ab})\,,
\end{equation}
where $\lambda_{ab}=\hbar/\sqrt{2 m_{ab}k_BT}$ is De Broglie wave length of relative motion, $m_{ab}^{-1}=m_a^{-1}+m_{b}^{-1}$, $a=e,i$.  In Eq.(\ref{Kelbg-pot})
\begin{equation}\label{Fx}
F(x) = 1- \exp\left(-x^2\right) +\sqrt{\pi}  x \left(1-
\mbox{erf}\left( x \right) \right) \quad.
\end{equation}
Another analytic approximation for the exact two-particle effective potential is the expression
derived by Deutsch which was used in the simulations by Hansen and McDonald 
\cite{hansen2}. 

The Kelbg potential is a good approximation for the two-particle Slater sum in the case of small parameters $\xi_{ab} = - (e_a e_b)/(k_BT \lambda_{ab})$ if the interparticle distance $r$ is sufficiently large.  However, at small interparticle distances it shows a deviation from the exact value of $- k_BT \cdot \ln ( S_{ab}(r=0))$. In order to describe the right behavior also at small distances it is better to use a corrected Kelbg potential defined by \cite{VE99}

\begin{equation} \label{corr-Kelbg}
u_{ab}(r) =
\left({e_a e_b}/{r}\right)\cdot \left\{ F (r / \lambda_{ab})-
r \frac{k_BT}{e_a e_b} 
\tilde{A}_{ab}(\xi_{ab}) \exp
\left(-(r / \lambda_{ab})^2 \right) \right\} \quad.
\end{equation}

In Eq. (\ref{corr-Kelbg}) the coefficient
$A_{ab}(T)$
is adapted in such a way that ${S_{ab}(r=0)}$ and his first derivative ${S'_{ab}(r=0)}$ have the exact value
corresponding to the two-particle wave functions of the free states \cite{gruenes-buch,VE99,Rohde}. The corresponding coefficients for the elctron-electron and for the electron-ion interaction read
\BEQ{Aee}
\tilde{A}_{ee}&=&\sqrt{\pi} |\xi_{ee}| +\ln \left[2 \sqrt{\pi} |\xi_{ee}| \, \int \frac{dy\,y \exp \left(-y^2 \right)}{\exp\left( \pi |\xi_{ee}|/y \right)-1} \right]
\\
\tilde{A}_{ei}&=&- \sqrt{\pi}\xi_{ei}+\ln \left[ \sqrt{\pi} 
\xi_{ie}^3 
\left( \zeta(3) + \frac{1}{4} \zeta(5) \xi_{ie}^2 \right) \right.
+ \left. 4 \sqrt{\pi} \xi_{ei} \, \int \frac{dy\,y \exp \left(-y^2 \right)}{1-\exp{\left(-\pi \xi_{ei}/y \right)}} \right]
\EEQ

We mention that in the region of high temperatures
\begin{equation}
T_r = T/T_I = \left({2 k_B T \hbar^2}/{m_{ie} e^4}\right) > 0.3 \quad.
\end{equation}
the Kelbg potential ($A_{ab}=0$) almost coincide with the corrected Kelbg potential Eq. (\ref{corr-Kelbg}). 
In the region of intermediate temperatures $0.1 < T_r < 0.3$ the Kelbg potential does not give a correct description of the two-particle Slater sum at short distances. Instead
we may use the corrected Kelbg-potential Eq.(\ref{corr-Kelbg}) to get an appropriate approximation for the Slater sum at arbitrary distances. 

The effective potentials derived from perturbation theory do not include bound 
state effects. The other limiting case of large $\xi_{ab}$ or small temperature $T_r < 0.1$, where bound states are of importance, can be treated by another approach \cite{ebeling}. Here a transition to the chemical picture is made, i.e. bound and free states have to be separated.

In the present work
we are interested in the regime of intermediate temperatures. In this regime the simulations of the dynamics may be performed with the potential Eq.(\ref{corr-Kelbg}).


\section{Force-force autocorrelation function} \label{FF}

The system under consideration is a two-component plasma consisting of electrons and ions which is described by the semiclassical model introduced in Sec \ref{model}. Let us choose the position of one of the charged particles (for example an electron) as a reference point. Hereafter we call this particle the first one. The semiclassical force acting on the first particle equals
\BE{force}
\DK{F}=-\DK{\Delta}_1 \sum_{j=2}^{N} u_{1j}(\DK{r}_1-\DK{r}_j)\,
\EE
$u_{ij}$ being the effective pair potential between the $i$th and $j$th particles, defined in Eq. (\ref{corr-Kelbg}).

Define now two functions characterizing the dynamics of the first particle. The first one
\BE{vacf}
C(t)=\frac{<\DK{v}(t) \cdot \DK{v}(0)>}{<v^2>}
\EE
is the velocity-velocity autocorrelation function (velocity acf), the second function
\BE{facf}
C(t)=\frac{<\DK{F}(t) \cdot \DK{F}(0)>}{<F^2>}
\EE
is the force-force autocorrelation function (force acf). In the above equations the brackets $<\ldots>$ denote averaging over the equilibrium ensemble of the semiclassical system. The velocity acf is formally a function expressing the single particle properties. However, it is connected with the force acf which involves the collective properties by the relation
\BE{vfrel}
\frac{\partial^2 C(t)}{\partial t^2} + \omega_1^2 D(t) =0 \,,
\EE
where $\om_1^2={<F^2>}/{3mk_B T}$.

Define the one-side Fourier transform of the velocity and force acf,
\BE{Fourier}
\hat{C}(\omega)=\int_0^{\infty}dt e^{i\omega t} C(t)\,,~\hat{D}(\omega)=\int_0^{\infty}dt e^{i\omega t} D(t) \,.
\EE
The Fourier transform of Eq.(\ref{vfrel}) reads
\BE{vfrelF}
 \hat{D}(\omega)= \frac{\omega^2 \hat{C}(\omega) - i \om}{\omega_1^2}\,.
\EE

In order to construct the both autocorrelation functions it is useful to consider the frequency moments of the real part of the velocity acf Fourier transform 
\BE{moments}
M_n=\frac{1}{2 \pi} \int_{-\infty}^{\infty} \om^n \hat{C}_r(\om) e^{-i \om t} d \om\;,~n=0,1,2,\ldots \,.
\EE
The zeroth moment is the initial value of the velocity acf,
\BE{C0}
M_0=C(0)=1\,.
\EE
Due to the parity of the function $\hat{C}_r(\om)$, all moments with odd numbers are equal to zero.

The second moment is expressable through the initial value of the force acf, 
\BE{C2}
M_2=\frac{1}{2 \pi} \int_{-\infty}^{\infty} \om^2 \hat{C}_r(\om) e^{-i \om t} d \om=\om_1^2 {D}(0)=\om_1^2 \,.
\EE
The fourth moment includes the correlation function of the time derivative of the force,
\BE{C4}
M_4=\frac{1}{2 \pi} \int_{-\infty}^{\infty} \om^4 \hat{C}_r(\om) e^{-i \om t} d \om=\om_1^2 \om_2^2\,,
\EE
where we have introduced the magnitude $\om_2^2={<\dot{F}^2>}/{<F^2>}$.

The Nevanlinna formula of the classical theory of moments \cite{AMT85,OT92} expresses the velocity acf Fourier transform
\BE{nev}
\frac{1}{\pi}\int_{_\infty}^{\infty} \frac{\hat{C}_r(\om)}{z-\om} d \om=-i\hat{C}(z)=\frac{E_{n+1}(z)+q_n(z)E_n(z)}{D_{n+1}(z)+q_n(z)D_n(z)}
\EE
in terms of a function $q_n=q_n(z)$ analytic in the upper half-plane ${\rm Im}\,z>0$ and having a positive imaginary part there ${\rm Im}\,q_n(\om+i\eta)>0,\,\eta>0$, it also should satisfy the limiting condition: $\left( q_n(z)/z \right) \to 0$ as $z \to \infty$ within the sector $\theta < {\rm arg}(z)<\pi-\theta$. In Eq.(\ref{nev}) we have employed the Kramers-Kronig relation connecting the real and imaginary part of $\hat{C}(\om)$. The polynomials $D_n$ (and $E_n$) can be found in terms of the first $2n$ moments as a result of the Schmidt orthogonalization procedure. The first orthogonal polynomials read 
\BEQ{poly}
D_1&=&z\,,~~D_2=z^2-\om_1^2\,,~~D_3=z(z^2-\om_2^2)\,,\\
E_1&=&1\,,~~E_2=z\,,~~~~~~E_3=z^2+\om_1^2-\om_2^2)\,.
\EEQ

Consider first the approximation $n=1$ leading to the correct frequency moments $M_0$ and $M_2$. Using the Nevanlinna formula and Eq. (\ref{vfrelF}) we obtain
\BE{CDn=1}
\hat{C}(z)=i \frac{z+q_1(z)}{z^2-\om_1^2+q_1 z}\,,~~~~~~\hat{D}(z)=i \frac{z}{z^2-\om_1^2+q_1 z}\,.
\EE
The physical meaning of the function $q_1(z)$ is that of a memory function \cite{BDCST96} since the inverse Fourier transform of Eq. (\ref{CDn=1}) is
\BE{memory}
\frac{\partial^2 C(t)}{\partial t^2}+\om_1^2C(t)+\int_0^t ds\, q_1(t-s) \frac{\partial C(s)}{\partial s}=0 \,.
\EE
We have no phenomenological basis for the choice of that function $q_1(z)$ which would provide the exact expression for $\hat{C}(z)$ and $\hat{D}(z)$. A simple approximation is to put the function $q_1(z)$ equal to its static value
\BE{static}
q_1(z)=q_1(0)=i \nu
\EE
and Eq. (\ref{memory}) simplifies to the equation of a damped oscillator with frequency $\om_1$ and damping constant $\nu$.
\BE{damposc}
\frac{\partial^2 C(t)}{\partial t^2}+\om_1^2C(t)+\nu \frac{\partial C(t)}{\partial t}=0 \,.
\EE

The static value $q_1(z=0)$ is connected with the self-diffusion coefficient $D$. The latter is defined by the time integral of the velocity acf
\BE{diff}
D = \frac{1}{\beta m_1}\int_0^{\infty} dt C(t) = \frac{1}{\beta m_1} \hat{C}(0)\,,
\EE
where $\beta=1/(k_B T)$ and $m_1$ is the mass of the first particle. With the use of Eqs. (\ref{diff}) and (\ref{static}) we obtain from Eq. (\ref{nev}) that $\nu = \om_1^2 \beta m_1 D$.

The inverse Fourier transform of Eq. (\ref{nev}) with the static approximation 
Eq. (\ref{static}) expresses the velocity and force acf's as a linear combination of two exponential functions $\exp(z_1t)$ and $\exp(z_2t)$, where $z_{1/2}=-{\nu}/{2} \pm \sqrt{\nu^2-4\om_1^2}/2$. Within this approximation we may distinguish between two regimes. In the first regime - the ``diffusion-regime'' - one deals with a large diffusion constant. As a result $\nu=\beta m_1 D \om_1^2>2\om_1$ and Eq. (\ref{damposc}) is the equation of an overdamped oscillator. In this regime the velocity autocorrelation function goes monotoneously to zero. With decreasing diffusion constant the damping constant $\nu$ becomes smaller. At certain thermodynamical conditions just the opposite inequality $\nu<2\om_1$ holds. This corresponds to an ``oscillatory-regime'' and at least one of the autocorrelation functions should show an oscillatory behavior. The existence of the two regimes have been established for the case of an OCP \cite{BDCST96} and has been confirmed  by our molecular-dynamics simulation for the case of a TCP. To obtain not only a qualitative  but also a quantitative correspondence with the results of MD simulations one has to go beyond the simple approximation  $n=1$ in the Nevanlinna formula Eq. (\ref{nev}).

Consider therefore the case $n=2$ in Eq. (\ref{nev}). Then the autocorrelation functions are expressed via the function $q_2(z)$ as
\BE{CDn=2}
\hat{C}(z)=i \frac{z^2+\om_1^2-\om_2^2+q_2(z)z}{z(z^2-\om_2^2)+q_2(z^2-\om_1^2)}\,,~~~~~~\hat{D}(z)=i \frac{z(z+q_2)}{z(z^2-\om_2^2)+q_2(z^2-\om_1^2)}\,.
\EE
Eq. (\ref{CDn=2}) reproduces the exact freqency moments from $M_0$ up to $M_4$. For the function $q_2(z)$ we choose again a static approximation
\BE{static2}
q_2(z) \equiv q_2(0) \equiv i h\,,
\EE
where $h$ has to be taken from the relation
\BE{hf}
h=\left(\frac{\om_2^2}{\om_1^2}-1 \right)/\beta m_1 D 
\EE
in order to obtain the exact low frequency value $\hat{C}(0)$ given by Eq. (\ref{diff}). 

From Eq. (\ref{CDn=2}) we find that the autocorrelation functions are now given by the linear combination of three exponentials,
\BE{3exp}
C(t)=\sum_{i=1}^3 C_i e^{i \Omega_i t}\,,~~~D(t)=\sum_{i=1}^3 d_i e^{i \Omega_i t}\,.
\EE
The complex frequencies $\Omega_i$ are the poles of the expressions Eq. (\ref{CDn=2}). They are defined as the solutions of the cubic equation,
\BE{cubic}
\Omega(\Omega^2-\om_2^2)+ih(\Omega^2-\om_1^2)=0\,.
\EE
The coefficients $C_i$ ($d_i$) characterizes the strength of the $i$th mode,
\BEQ{Ci,Di}
C_i&=&\frac{\om_1^2}{\Omega_i^2}d_i\,,~i=1,2,3\,,\\
d_1&=&i(h+i\Omega_1)\Omega_1 (\Omega_2-\Omega_3)/N\,\\
d_2&=&i(h+i\Omega_2)\Omega_2 (\Omega_3-\Omega_1)/N\,\\
d_3&=&i(h+i\Omega_3)\Omega_3 (\Omega_1-\Omega_2)/N\,\\
N&=&(\Omega_1-\Omega_2)(\Omega_3-\Omega_1)(\Omega_2-\Omega_3) \,.
\EEQ

Equations (\ref{3exp}) constitute the basic approximation of our paper. The frequencies $\Omega_i$ and the  coefficients $C_i$ (or $d_i$, respectively) are expressed by three parameters - the diffusion constant $D$, and the frequencies $\om_1$ and $\om_2$. The constructed autocorrelation functions satisfy the following conditions: (i) the exact short time behavior for the velocity acf is reproduced to the orders $t^2$ and $t^4$, (ii) the short time behavior of the force acf is reproduced to the order $t^2$, (iii) the long time behavior of the velocity acf generates the exact diffusion constant, and (iv) the connection between the velocity and force acf's Eq. (\ref{vfrel}) is satisfied.

The parameters $D$, $\om_1$ and $\om_2$ may be calculated by another approximations. The both frequencies $\om_1$ and $\om_2$ are expressable via the partial correlation functions of our semiclassical system. 
The parameter $\om_1$ is given by the electron-ion and electron-electron partial pair correlation functions. To calculate the frequency $\om_2$ one needs the knowledge of the partial ternary distribution functions. The diffusion constant may be obtained from kinetic theory. In contrast to the case of an OCP \cite{BDCST96} the parameters to be calculated are very sensitive to the approximations used to calculate the static distribution functions. Therefore in this paper we take the ``input'' parameters directly from the computer simulations.


To check the quality of the predictions from our approximation we have performed molecular dynamics simulations for comparison. The equations of motions obtained with the effective potential Eq.(\ref{corr-Kelbg}) were integrated numerically for the case of equal masses $m_e=m_i$ using the leap-frog
variant of Verlet's algorithm. 
The simulations were performed for 128 electrons and 128 positrons moving in a cubic box with  periodic
boundary conditions. In the investigated range of plasma parameters ($T=30\,000~{\rm K}$, the coupling parameter has varied from $\Gamma =0.2$ up to $\Gamma =3$) the size of the simulation box was significantly greater than the Debye radius. Therefore the long-range Coulomb forces are screened inside each box and there was no need to use the Ewald summation instead the simple periodic boundary conditions.  The thermal
equilibrium in the system was established (and maintained) by a Langevin source. Such simulations has been recently used to obtain the static distribution of the electric microfield at a charged particle \cite{OVE99}. In this paper we extract the velocity and force autocorrelation functions as the main characteristics of the microfield dynamics. 

\begin{table}[tb] \label{tab1} {TABLE I. The $\Gamma$ dependence of the parameters $\om_1$, $\om_2$ and $D$. $\om_1$ and $\om_2$ are given in units of electron plasma frequency $\om_{pe}=\sqrt{4 \pi n_e e^2/m_e}$, $D$ is given in units of $1/(m_e \om_{pe} \beta)$}\\[1ex]
\begin{tabular}{|c||c|c|c|} 
$\Gamma$ & $\om_1$ & $\om_2$& $D$ \\ \hline
0.2&0.84&13.6&10.3\\
1.5&0.88&3.3&4.41\\
3.0&0.61&2.1&5.75 
\end{tabular}
\end{table}

In Figs. 1-3 we present the results of the MD data. 
The simulation results  are compared with our analytical approximation Eqs. (\ref{CDn=2}). The three input parameters for the analytical approximation are taken from the MD simulations. The diffusion constant is obtained from the time integral of the velocity acf (Eq. (\ref{diff})). Since the velocity acf is a slowly decaying function it requires a long simulation time to extract the diffusion constant. For our model system with equal electron and ion masses it is possible to perform the necessary simulations. The frequency $\om_2$ has been taken from the exact short time behavior of the force acf $D(t)=1-\om_2^2 t^2/2$. Finally the frequency $\om_1$ was choosen to fit the model to the data. In Table I we show the parameters $\om_1$, $\om_2$ and $D$ for three coupling parameters $\Gamma$ considered in this paper.

Except the case of the force acf at $\Gamma=0.2$ there is a good overall agreement between the theoretical approximations and the MD data.  We believe that the strong deviation of the MD data from the theoretical predictions for  $\Gamma=0.2$ is a numerical artefact due to the poor statistics in the weak coupling case. From the figures we see that with increasing plasma parameter $\Gamma$ the dynamics of the charged particles switches from the diffusion-like regime at $\Gamma=0.2$ to the oscillator-like motion at $\Gamma=3.0$.  The value $\Gamma=1.5$ may be considered as a critical value separating the both regimes. We may also see from the figures that the oscillator-like motion is more pronounced for the force acf.

 At still higher densities ($\Gamma \ge 3$ at $T=30\,000~{\rm K}$) the semi-classical approach employed in this paper fails to describe the quantum two-component plasma properly.

\section{Conclusions}

The electric microfield dynamics at a charged particle in a two-component plasma has been studied. The quantum plasma has been modeled by a semiclassical system with effective potentials.  The effective potential was choosen to describe the nondegenrate limit of the quantum system appropriately. We have investigated the velocity and force acf's of the semiclassical system. The starting point for the theoretical analysis was the exact expression of the autocorrelation functions through the Nevanlinna formula Eq. (\ref{CDn=2}), satisfying three sum rules for the velocity acf. The approximation Eq. (\ref{static2}) together with Eq. (\ref{hf}) expresses the velocity acf in terms of three parameters. Two of them - $\om_1$ and $\om_2$ - describe the exact short time behavior of the velocity acf up to the order $t^4$, the third parameter, the self-diffusion constant $D$ is related to the time integral of the velocity acf. Since the force acf can be obtained from the velocity acf by a second time derivative the force acf is expressed through the same three parameters. The general picture is as follows. At weak coupling the diffusion of the charged particle dominates the collective plasma oscillations and the particle motion is diffusion-like. In this regime the velocity acf decays exponentially with a decay rate $1/D$ (time in units of the inverse electron plasma frequency $\om_{pe}$). The force acf has a positive decay at short times (decay rate $\om_1^2 D$) and a negative decay at long times (with the rate $1/D$). At strong coupling the diffusion is supressed and a weakly damped oscillatory behavior for the force acf developes. 
The theoretical predictions has been compared with molecular dynamics simulations data.
There is an overall agreement of the force dynamics obtained by the analytical approximation with the MD data.

Finally, we mention that there is no one to one correspondence of the semiclassical autocorrelation functions with the corresponding characteristics of the quantum system. Nevertheless, we suspect that the semiclassical force dynamics considered in this paper at least qualitatively reproduces the electric microfield dynamics of the quantum system. 

{\bf Acknowledgments.} This work was supported by the Deutsche Forschungsgemeinschaft (DFG) and the Deutscher Akademischer Austauschdienst (DAAD) of Germany.



\newpage

\begin{center}
{\bf FIGURE CAPTIONS}
\end{center}

\begin{description}

\item[(Figure 1)] Time dependence of velocity acf $C(t)$ and force acf $D(t)$ at $\Gamma=0.2$. Time is in units of inverse electron plasmafrequency $\om_{pe}^{-1}$. Solid lines: present theoretical approximation; Points: results of molecular-dynamics simulations.

\item[(Figure 2)] Same as in Fig. 1 at $\Gamma=1.5$.

\item[(Figure 3)] Same as in Fig. 1 at $\Gamma=3.0$.

\end{description}

\newpage

\vspace*{-2cm}
\begin{figure}[h]
\unitlength1mm
  \begin{picture}(155,150)
\put (0,0){\psfig{figure=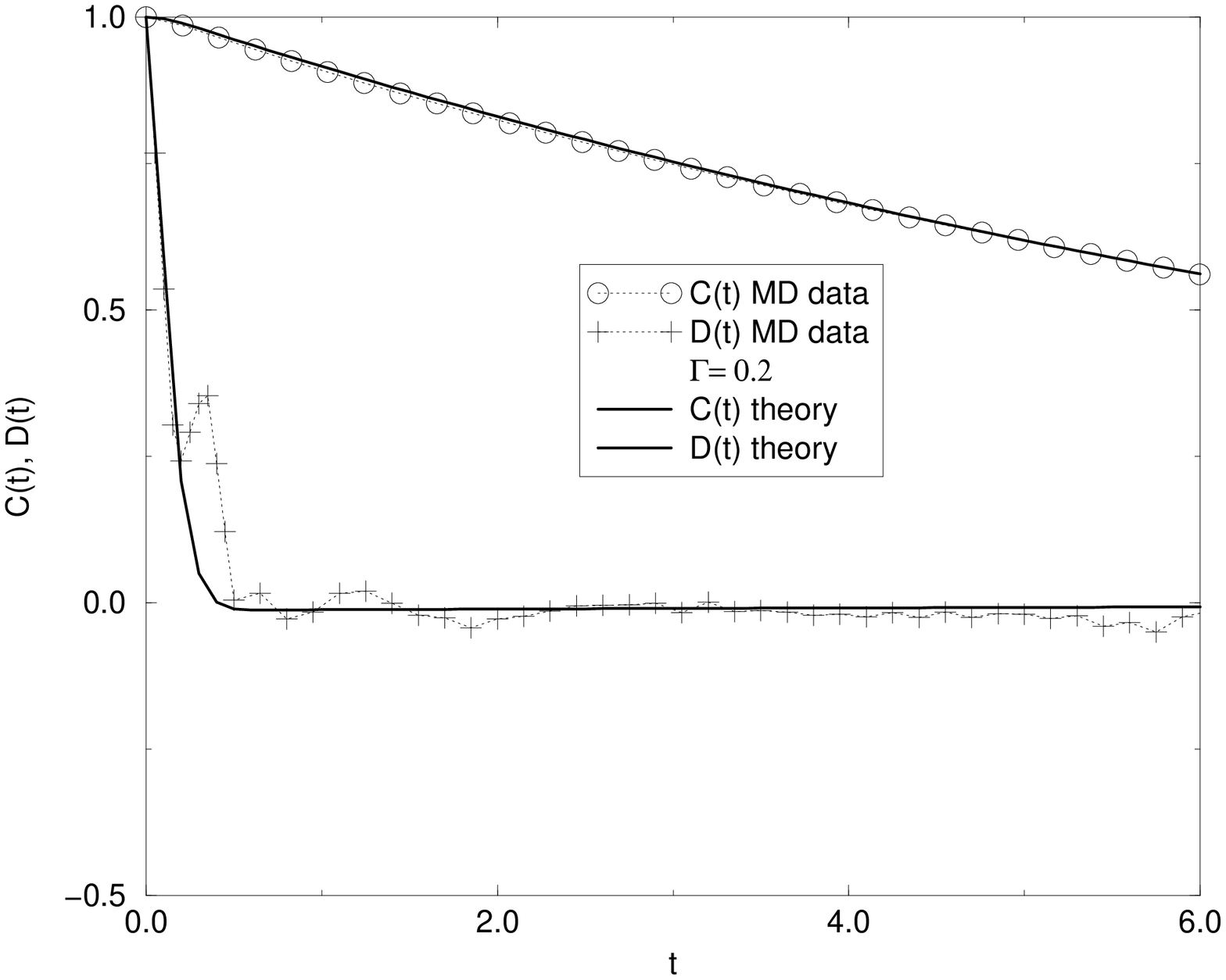,width=15.0cm,height=14.0cm,angle=-90}}
 \end{picture}\par
\end{figure}

\newpage

\vspace*{-2cm}
\begin{figure}[h]
\unitlength1mm
  \begin{picture}(155,150)
\put (0,0){\psfig{figure=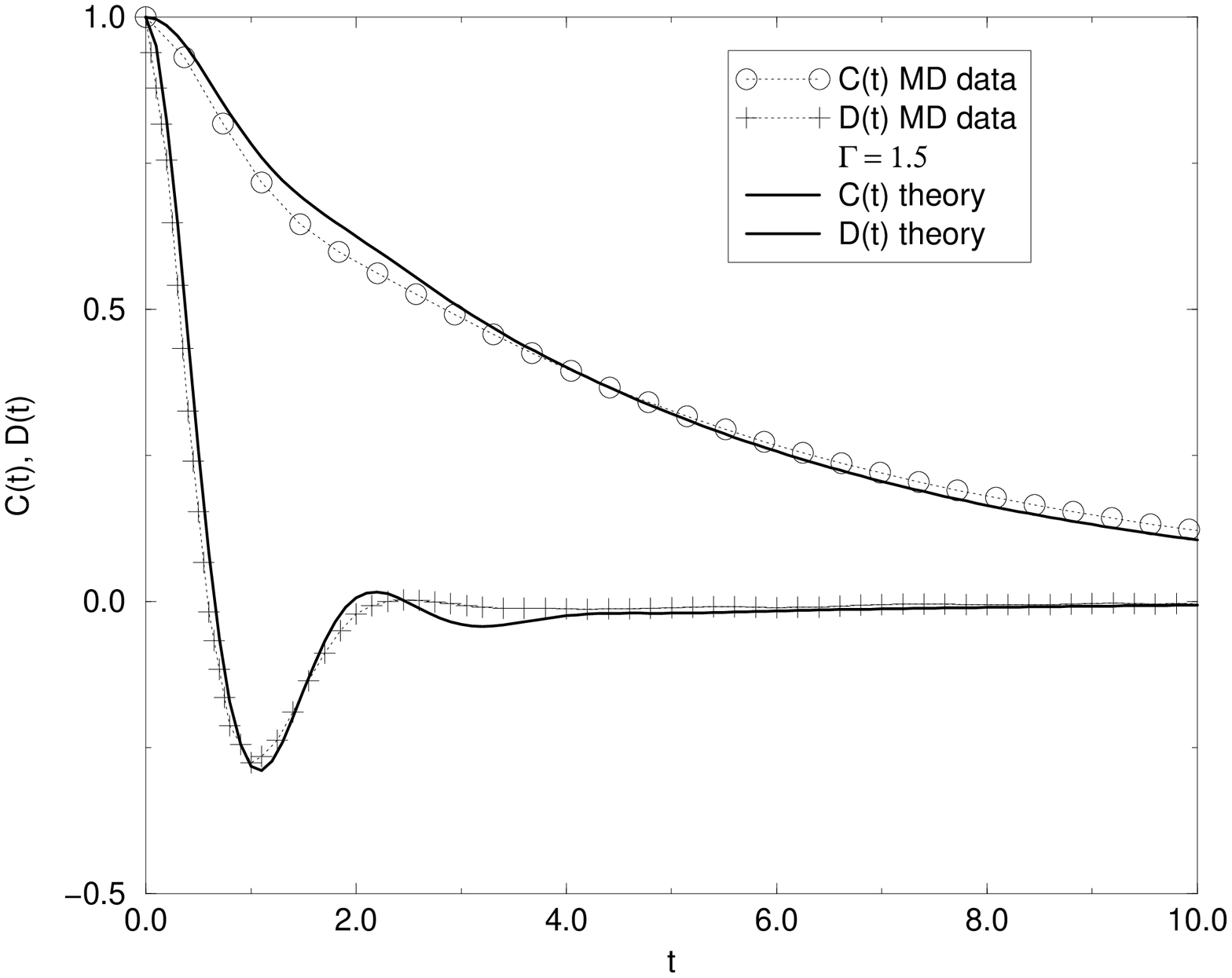,width=15.0cm,height=14.0cm,angle=-90}}
 \end{picture}\par
\end{figure}

\newpage

\vspace*{-2cm}
\begin{figure}[h]
\unitlength1mm
  \begin{picture}(155,150)
\put (0,0){\psfig{figure=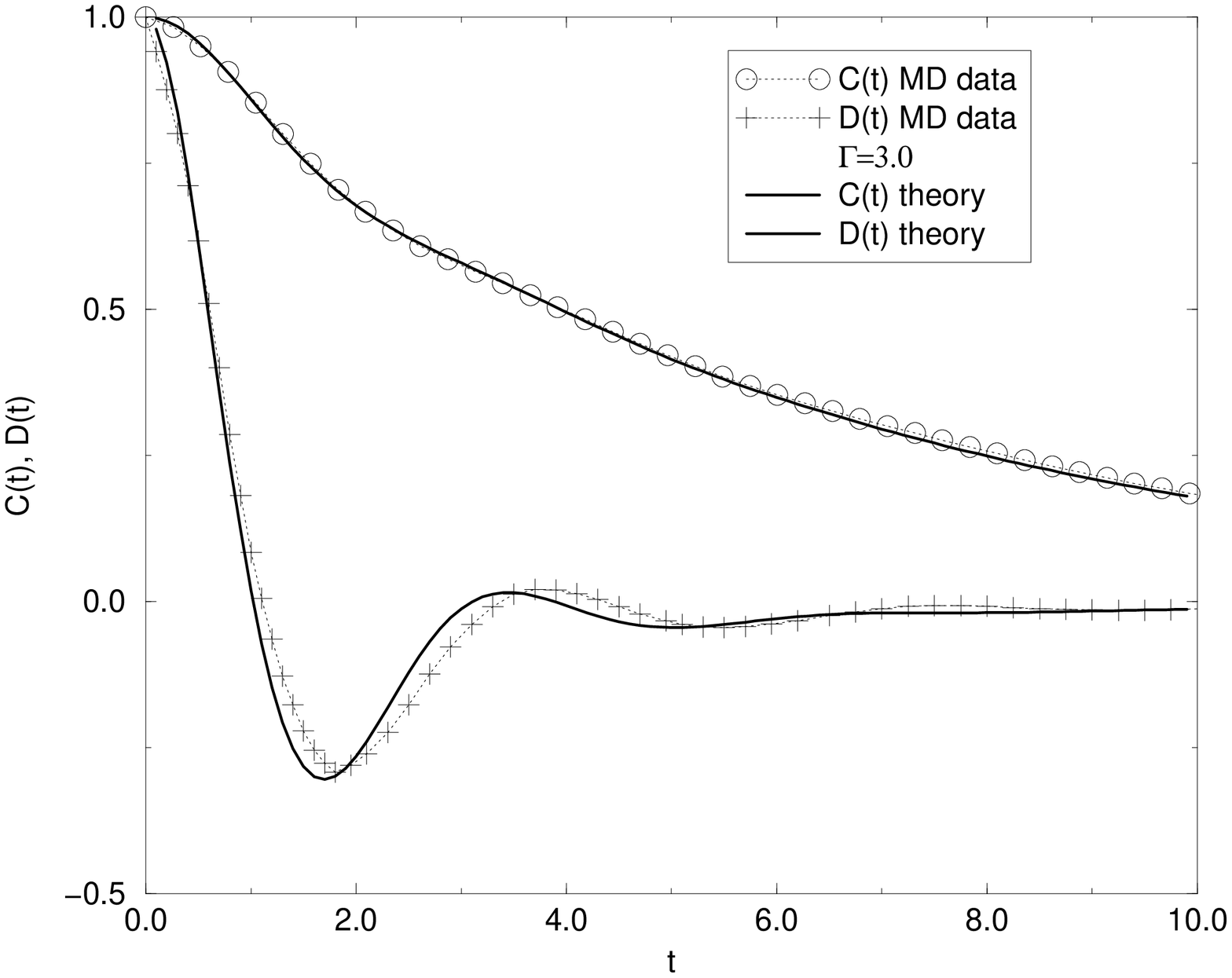,width=15.0cm,height=14.0cm,angle=-90}}
 \end{picture}\par
\end{figure}

\end{document}